\documentclass[12pt]{article}
\usepackage[colorlinks, citecolor=blue,anchorcolor=red,menucolor=red, linkcolor=red,filecolor=red,urlcolor=blue,frenchlinks=red]{hyperref}
\usepackage{multirow}
\usepackage[left=2cm,right=2cm,top=2cm,bottom=2cm]{geometry}
\usepackage{cite}
\usepackage{psfrag}
\usepackage[demo]{graphicx}
\usepackage{graphicx}
\usepackage{graphics}
\usepackage{epsfig}
\usepackage{color}
\usepackage{booktabs}
\usepackage{amsmath,amsfonts,amssymb}
\usepackage{pstcol,pst-fill,pst-grad}
\usepackage{pstricks,pst-fill,pst-grad}
\usepackage{euscript}
\usepackage{pstricks}
\usepackage{wrapfig}
\usepackage{slashed}
\usepackage[toc,page]{appendix}
\usepackage{here}

\begin{document}
	\title{\vspace{-3cm}
		\hfill\parbox{4cm}{\normalsize \emph{}}\\
		\vspace{1cm}
		{}}
	\vspace{2cm}
	\author{\thanks{} \\
		{\it {\small}}\\
		{\it {\small }}\\
		{\it {\small}}\\				
	}
	\title{\vspace{-3cm}
		\vspace{1cm}
		{Laser-assisted doubly charged Higgs pair production in Higgs triplet model (HTM)}}
		\vspace{2cm}
	\author{J. Ou aali,$^1$ M. Ouali,$^1$ M. Ouhammou,$^1$ S. Taj,$^1$ B. Manaut$^{1}$\thanks{Corresponding author, E-mail: b.manaut@usms.ma} and L. Rahili$^2$  \\
		{\it {\small$^1$ Polydisciplinary Faculty, Laboratory of Research in Physics and Engineering Sciences,}}\\
		{\it {\small Team of Modern and Applied Physics, Sultan Moulay Slimane University, Beni Mellal 23000, Morocco.}}\\
		{\it {\small$^2$ EPTHE, Faculty of Sciences, Ibn Zohr University, B.P 8106, Agadir, Morocco.}}\\			
	}
	
	\maketitle \setcounter{page}{1}
\date{}
\begin{abstract}
In the framework of Higgs triplet model (HTM), we study the pair production process of doubly charged Higgs bosons via $e^{+}e^{-}$ annihilation in the presence of a laser field with circular polarization . 
We begin our work by presenting the theoretical calculation of the differential cross section in the centre of mass frame including both $Z$ and $\gamma$ diagrams. 
Then, from the numerical analysis of the production cross section's dependence on the laser field parameters, we have shown that the laser-assisted total cross section decreases as far as the  electromagnetic field intensity enhances or by decreasing its frequency. 
Finally, we analyze the variation of the total cross section versus the mass of the doubly charged Higgs boson by fixing the laser field parameters and the centre of mass energy, and we have found that the order of magnitude of the cross section decreases as long as $M_{H^{\pm\pm}}$ increases. 
\end{abstract}
\maketitle
\section{Introduction}
After the discovery of the laser technology in 1960 \cite{1}, the study of laser-assisted processes has been an active area of both theoretical \cite{2,3} and experimental \cite{4,5} research. This is due not only to its fundamental importance in collision physics but also to its role in assisting our understanding and interpreting a wide range of scientific phenomena and technological applications \cite{6,7}.
The study of laser-assisted processes provides important information about behavior of particles and their properties.
In this respect, a significant deal of theoretical works has been consecrated to the study of weak decay and scattering processes \cite{8,9}, and it is found that the circularly polarized laser field prolongs the particle's lifetime and enhances its modes decay.
In the past few years, there has been particular interest in the study of the electron-positron interactions in the presence of an external field. Moreover, the effect of the electromagnetic field with circular polarization is that it decreases the total cross section.
Detailed reports on laser-assisted elementary particles' production via electron-positron annihilation can be found in the papers \cite{10,11}. 
In addition, the laser-assisted electron-positron annihilation allows the observation of a variety of new phenomena related to high energy physics such as the standard model of particle physics and beyond. 
In ref \cite{12}, we have found that the electromagnetic field reduces the total cross section of the production of a charged Higgs boson in association with a charged weak W-boson ($H^{\pm}W^{\mp}$).

The LHC milestone discovery \cite{13,14} of the Higgs boson in July 2012 with a mass 125 GeV has been a great mutation in modern particle physics because it provides us with a deep understanding of the electroweak symmetry breaking mechanism (EWSB) which is outlined in the Standard Model of particle physics by Brout-Englert Higgs mechanism \cite{15,16}.
After some years of accumulated data, the properties of this new particle are in excellent agreement with the prediction of the standard model \cite{17}. 
However, this current experimental result raises the question of whether there are other fundamental scalars that may solve some of the open questions in particle physics such as the origin of neutrino mass and the existence of dark matter. 
An alternative philosophy is based on adding the minimal required number of new fields to the Standard Model in order to address these problems.
One example is the SM extended by two real scalar fields which can provide a viable dark matter candidate \cite{18,19}, and it is also very rich in terms of its collider phenomenology \cite{20}.

The Higgs triplet model (HTM) \cite{21,22} is one of the simplest models beyond the SM which postulates that the SM Higgs sector should be extended by a triplet field $\Delta$ with $Y = 1$. 
The spectrum of the HTM contains two doubly charged $H^{\pm\pm}$, two singly charged $H^{\pm}$, one CP-odd $A^{0}$ and two CP-even $h^{0}$ and $H^{0}$. 
The characteristic feature of such model with Higgs triplet is that it proposes the most direct way to explain neutrino masses \cite{23,24}. 
In addition, it is well known that the doubly charged Higgs bosons $H^{\pm\pm}$ can be seen as the typical particles in this model. 
In the HTM, there are two main decay modes for $H^{\pm\pm}$: the same-sign diboson decay ($H^{\pm\pm}\rightarrow W^{\pm}W^{\pm}$) and the same-sign dilepton decay ($H^{\pm\pm}\rightarrow l^{\pm}l^{\pm}$). 
The collider phenomenology of a doubly charged scalar has been discussed in \cite{25}. 
So that, if a doubly charged Higgs boson is discovered at LHC, it will be critical to determine its couplings at the future high-energy linear colliders such as the Compact Linear Collider (CLIC) \cite{26} and the International Linear Collider (ILC) \cite{27}, and this is due to their very clean environment and high luminosity. The pair production of doubly charged Higgs boson at $e^{+}e^{-}$ colliders has been studied in Refs \cite{28,29}. 
In this respect, the main purpose of this paper is to study the production process $e^{+}e^{-}\rightarrow H^{++}H^{--}$ inside a circularly polarized electromagnetic field to analyze theoretically and analytically its effect on the production cross section in HTM model.

Our work is organized as follows: In section \ref{sec2}, we briefly review the model under consideration. Then, we give the relevant couplings which are related to our calculation as well as the theoretical calculation of the total cross section of doubly charged Higgs production through $e^{+}e^{-}$ annihilation in the presence of a circularly polarized laser field. In section \ref{sec3}, we discuss the laser-assisted production cross-sections for the process $e^{+}e^{-}\rightarrow H^{++}H^{--}$, and we present some phenomenological analysis. Finally, we give our conclusion in section \ref{concl} We mention that, in this paper, we have used natural units $\hbar=c=1$, and the metric tensor $g^{\mu\nu}$ is taken such as $g^{\mu\nu}=(1,-1,-1,-1)$.
\section{Outline of the theory}\label{sec2}
\subsection{Scalar potential $\pmb{\&}$ Relevant Couplings }
The Higgs Triplet model is described in detail in Refs \cite{30,31}. Here, we only briefly review the theoretical setup as well as the main couplings relevant to the present work.
The scalar sector of HTM consists of the SM Higgs field, $H$, and a triplet scalar field $\Delta$ with hypercharge $Y=\, 1$, and they are given by:
\begin{equation}
H=\begin{pmatrix}
\Phi^{+} \\
\Phi^{0}
\end{pmatrix},  \quad\qquad \Delta=\begin{pmatrix}
\delta^{+}/\sqrt{2} & \delta^{++} \\
\delta^{0} & -\delta^{+}\sqrt{2}
\label{eq1}
\end{pmatrix}.
\end{equation}
The most general renormalizable and gauge invariant potential is given by \cite{32}:
\begin{equation}
\begin{split}
V(H,\Delta) =& -\mu_{H}^{2}H^{\dagger}H+\frac{\lambda}{4}(H^{\dagger}H)^{2}+\mu_{\Delta}^{2}Tr(\Delta^{\dagger}\Delta) \\
&+\lambda_{1}(H^{\dagger}H)Tr(\Delta^{\dagger}\Delta)+\lambda_{2}(Tr\Delta^{\dagger}\Delta)^{2}+\lambda_{3}Tr(\Delta^{\dagger}\Delta)^{2} \\
&+\lambda_{4}H^{\dagger}\Delta\Delta^{\dagger}H+[\mu(H^{T}i\tau_{2}\Delta^{\dagger}H) + h.c.]
\end{split}
\label{eq2}
\end{equation}
In the above potential, $\mu_{H}$ and $\mu_{\Delta}$ stand for mass squared parameters. $\lambda,\, \lambda_{i=1..4}$ are dimensionless couplings, while $\mu$ denotes dimensionless coupling that mixes two Higgs fields. After the electroweak symmetry breaking, the neutral component, $\Phi^{0}$ and $\delta^{0}$, can acquire vevs such that:
\begin{equation}
H=\frac{1}{\sqrt{2}}\begin{pmatrix}
0 \\
v_{\Phi}
\end{pmatrix},  \quad\qquad \Delta=\frac{1}{\sqrt{2}}\begin{pmatrix}
0 & 0 \\
v_{\Delta} & 0
\end{pmatrix},
\label{eq3}
\end{equation}
with $v^{2}=v_{\Phi}^{2}+2v_{\Delta}^{2}=(246\,  GeV)^{2}$. The spectrum of the scalar potential will have seven scalar particles. Indeed, in addition to the pair of doubly-charged Higgs bosons $H^{\pm\pm}$, the HTM provides a pair of charged Higgs bosons $H^{\pm}$ that appear together with the charged Goldstone $G^{\pm}$ after an orthogonal rotation by using the mixing matrix defined by $R_{\beta_{\pm}} = \lbrace\lbrace\cos\beta_{\pm},-\sin\beta_{\pm}\rbrace,\lbrace\sin\beta_{\pm},\cos\beta_{\pm}\rbrace\rbrace$ where $\beta_{\pm}$ denotes the angle between the non-physical fields $\Phi^{\pm}$ and $\delta^{\pm}$ such that $\tan\beta_{\pm}=\sqrt{2} v_{\Delta}/v_{\Phi}$.
Analogously, the two CP-even neutral scalars ($h^{0},H^{0}$) and the two CP-odd neutral pseudo-scalars ($G^{0}, A^{0}$) are obtained by an orthogonal transformation using the following two unitary rotations, $R_{\alpha}$ and $R_{\beta_{0}}$, respectively (see \cite{31} for more details). Moreover, in the HTM, the gauge couplings of the doubly charged scalars, which are related to our calculation, can be written as \cite{33,34}:
\begin{equation}
A^{\nu}H^{++}H^{--} = -2ie(p_{4}-p_{3})^{\nu},\quad\qquad Z^{\nu}H^{++}H^{--} = -i\frac{e(1-2S_{W}^{2})}{S_{W}C_{W}}(p_{4}-p_{3})^{\nu}.
\label{eq4}
\end{equation}
The short notations $C_W$ and $S_W$ represent successively $\cos({\theta_W})$ and $\sin({\theta_W})$, where $\theta_W$ is the Weinberg angle. $p_{3}$ and $p_{4}$ indicate the free four-momentum of $H^{--}$ and $H^{++}$, respectively.
\subsection{Laser-assisted cross section}
\begin{figure}[H]
  \centering
      \includegraphics[scale=0.33]{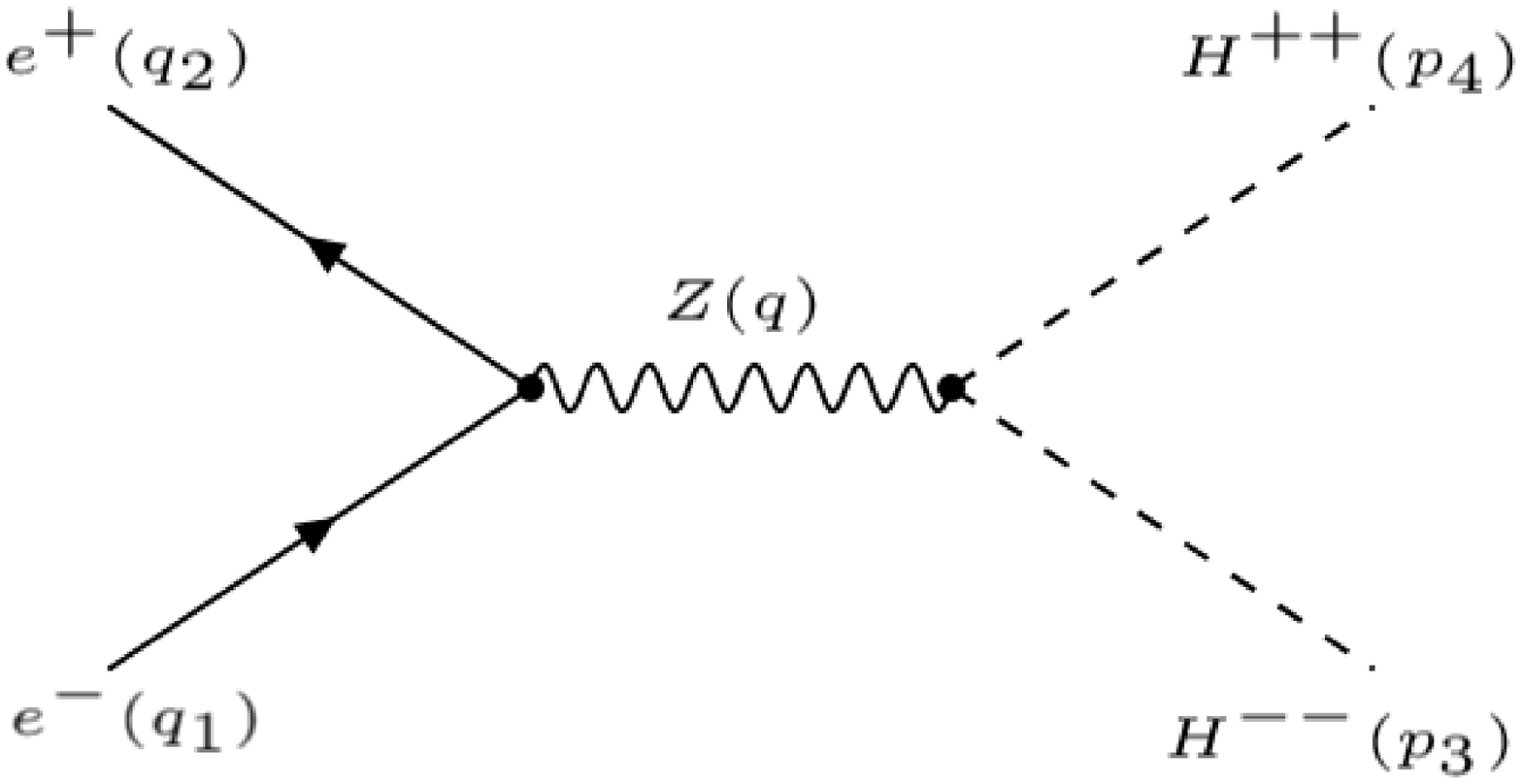}\hspace*{0.4cm}
      \includegraphics[scale=0.33]{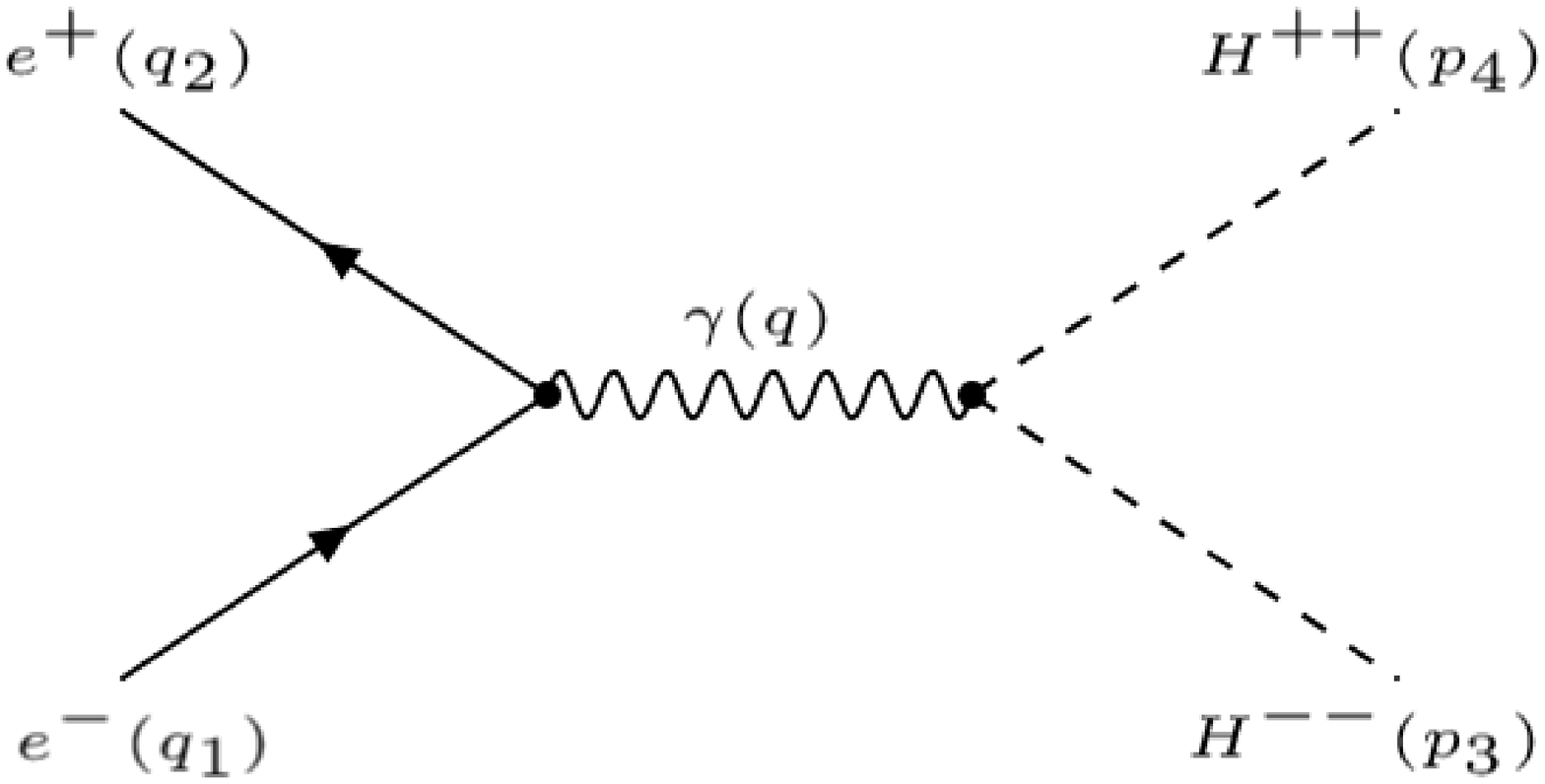}\par\vspace*{0.5cm}
        \caption{Leading-order Feynman diagrams of the process $e^{+}e^{-}\rightarrow H^{++}H^{--}$.}
        \label{fig1}       
\end{figure}

In this section, we perform a theoretical calculation of the differential cross section of the process $e^{+}e^{-}\rightarrow H^{++}H^{--}$ at tree-level in the presence of a laser field. The Feynman diagrams of this process are shown in figure \ref{fig1}.
In this study, we have considered the produced Higgs bosons $H^{\pm\pm}$ as free particles, while the electron and positron are embedded  in a laser field which is considered as a plane, monochromatic and circularly polarized electromagnetic wave, and it is given by the following classical four-potential:
\begin{equation}
A^{\mu} = \eta_{1}^{\mu}\cos\phi + \eta_{2}^{\mu}\sin\phi.
\label{eq5}
\end{equation}
Here, $\phi=k.x$ is the phase of the laser field. $k=(\omega,0,0,\omega)$ is the wave four vector, with $\omega$ is the laser frequency. $\eta_{1}=(0,\eta,0,0)$ and $\eta_{2}=(0,0,\eta,0)$ are the polarization four vectors, and they satisfy the following equations: ($\eta_{1}.\eta_{2}$)=0 and $\eta_{1}^{2}=\eta_{2}^{2}=\eta^{2}=-\lvert\pmb{\eta}\lvert^2=-(\varepsilon_{0}/\omega)^2$ where $\varepsilon_{0}$ is the electric field strength. The application of Lorentz condition, $\partial_{\mu}A^{\mu}=0$, implies that $k.\eta_{1} = 0$ and $k.\eta_{2} = 0$. In the presence of an electromagnetic field, the incident particles are described by Dirac-Volkov functions. The latter are written as follows:
\begin{equation}
\begin{cases}
\psi_{p_{1}, s_{1}}(x)= \Big[1-\dfrac{e \slashed k \slashed A}{2(k.p_{1})}\Big] \frac{u(p_{1}, s_{1})}{\sqrt{2Q_{1}V}} \exp^{iS(q_{1},s_{1})}&\\
\psi_{p_{2}, s_{2}}(x)= \Big[1+\dfrac{e \slashed k \slashed A}{2(k.p_{2})}\Big] \frac{v(p_{2}, s_{2})}{\sqrt{2Q_{2}V}} \exp^{iS(q_{2},s_{2})},
\end{cases}
\label{eq6}
\end{equation}
where the first term stands for the electron's state, and the second one represents the positron's state. $x$ indicates the space time coordinate of both the electron and positron. $u(p_{1}, s_{1})$ and $v(p_{2}, s_{2})$ are their Dirac bispinors where $s_i\, (i = 1, 2)$ denote their spins which satisfy the following sum-rules: $\sum_{s}u(p_{1}, s_{1})\bar{u}(p_{1},s_{1})=(\slashed p_{1}-m_{e})$ and $\sum_{s}v(p_{2}, s_{2})\bar{v}(p_{2}, s_{2})=(\slashed p_{2}+m_{e})$. $p_{1} = (E_{1},\lvert p_{1}\lvert,0,0)$ and $p_{2} = (E_{2},-\lvert p_{1}\lvert,0,0)$ are referring to their corresponding free four momentum in the centre of mass frame. In the presence of the electromagnetic field, the electron and positron get non-vanishing effective energy given by $Q_{i}\, (i= 1, 2)$. We define the arguments of the exponential terms in equation (\ref{eq6}) such as: 
\begin{equation}
\begin{cases}
S(q_{1},s_{1})=- q_{1}x +\frac{e(\eta_{1}.p_{1})}{k.p_{1}}\sin\phi - \frac{e(\eta_{2}.p_{1})}{k.p_{1}}\cos\phi&\\
S(q_{2},s_{2})=+ q_{2}x +\frac{e(\eta_{1}.p_{2})}{k.p_{2}}\sin\phi - \frac{e(\eta_{2}.p_{2})}{k.p_{2}}\cos\phi,
\end{cases}
\label{eq7}
\end{equation}
with $q_{i}\,(i=1, 2) = p_{i} + e^{2}a^{2}/2(k.p_{i})k$ represents the effective momentum of the electron and positron inside  the laser field such that:
\begin{equation}
q_{i}^{2}=m_{e}^{*2} = m_{e}^{2} + e^{2}a^{2}.
\label{eq8}
\end{equation}
Here, $m_{e}^{*}$ is the effective mass of the incident particles. $e$ is the charge of the electron, and $m_e$ denotes its mass. As mentioned before, the produced particles are free, then there is no interaction between the doubly charged Higgs Boson $H^{\pm\pm}$ and the laser field. Therefore, they are described by Klein-Gordon states, which are given by the following expressions:
\begin{equation}
\varphi_{p_{3}}(y)=\dfrac{1}{\sqrt{2 E_{H^{--}} V}} e^{-ip_{3}y} \hspace*{0.1cm},\hspace*{0.5cm}  \\\ \varphi_{p_{4}}(y)=\dfrac{1}{\sqrt{2 E_{H^{++}} V}} e^{-ip_{4}y},
\label{eq9}
\end{equation}
Here, $y$ is the space time coordinate of outgoing particles. $p_{3}$ = ($E_{H^{--}}$,$|p_3|\cos\theta$,$|p_3|\sin\theta$,0) and $p_4$ = ($E_{H^{++}}$,$-|p_4|\cos\theta$,$-|p_4|\sin\theta$,0) are the free four-momentum of the doubly charged Higgs-bosons, where $E_{H^{--}}$ and $E_{H^{++}}$  are their corresponding energies.
By using the gauge couplings given by equation (\ref{eq4}), the scattering-matix element \cite{35} of the laser-assisted doubly charged Higgs-boson pair production in HTM can be written as:
\begin{eqnarray}
S_{fi}({e}^{+}{e}^{-}\rightarrow H^{++}H^{--})&=&  \int_{}^{} d^4x  \int {}^{} d^4y \Big\lbrace \bar{\psi}_{p_{2}, s_{2}}(x)   \Big( \frac{-ie}{2C_{W} S_{W} } \gamma^{\mu} (g_v^{e} -g_a^{e}\gamma^{5})   \Big)   \psi_{p_{1}, s_{1}}(x) D_{\mu\nu}(x-y) \nonumber \\
 &\times & \varphi^{*}_{p_{3}}(y)(-i\frac{e(1-2S_{W}^{2})}{S_{W}C_{W}}(p_{4}-p_{3})^{\nu})\varphi^{*}_{p_4}(y)+\bar{\psi}_{p_{2}, s_{2}}(x)(-ie \gamma^{\mu})\psi_{p_{1}, s_{1}}(x) \nonumber \\ 
 &\times & G_{\mu\nu}(x-y)\varphi^{*}_{p_{3}}(y) (-2ie(p_{4}-p_{3})^{\nu})  \varphi^{*}_{p_4}(y)\Big\rbrace,
 \label{eq10}
\end{eqnarray}
where $g_v^{e}=-1 + 4\sin^{2}({\theta_W})$ and $g_a^{e}=1$ stand respectively for the vector and axial vector coupling constants. The factors $D_{\mu\nu}(x-y)$ and $G_{\mu\nu}(x-y)$ are the Feynman propagators for $Z$-boson and $\gamma$-boson, respectively. Their expressions are given by:
\begin{equation}
D_{\mu\nu}(x-y)=\int \dfrac{d^{4}q}{(2\pi)^4} \frac{e^{-iq(x-y)}}{q^{2}-M_{Z}^{2}}\Big (-ig_{\mu\nu}+i(1-\xi)\frac{q_{\mu}q_{\nu}}{M_{Z}^{2}}\Big),
\label{eq11}
\end{equation}
\begin{equation}
G_{\mu\nu}(x-y)=\int \dfrac{d^{4}q}{(2\pi)^4} \frac{e^{-iq(x-y)}}{q^{2}-M_{Z}^{2}}\Big (-ig_{\mu\nu}+i(1-\xi)\frac{q_{\mu}q_{\nu}}{q^{2}}\Big),
\label{eq12}
\end{equation}
with $\xi=1$ for the Feynman gauge or $\xi=0$ for the Lorentz gauge. $q$ indicates the four-momentum of the off-shell $V$ ($V=Z$ or $\gamma$). After introducing a short description of the procedure used to analyze the effects of laser field on the process \ref{fig1}, we substitute the expressions of wave functions (equations (\ref{eq6}) and (\ref{eq9})) and Feynman propagators (equations (\ref{eq11}) and (\ref{eq12})) into the equation (\ref{eq10}). Thus, we obtain:
\begin{eqnarray}
S_{fi}^{n}({e}^{+}{e}^{-}\rightarrow H^{++}H^{--})&=&\dfrac{(2\pi)^{4}\delta^{4}(p_{3}+p_{4} -q_{1}-q_{2}-nk)}{4V^{2}\sqrt{Q_{1}Q_{2}E_{H^{--}}E_{H^{++}}}} \big(M_{Z}^{n} + M_{\gamma}^{n} \big).
\label{eq13}  
\end{eqnarray}
In the above scattering matrix element expression (\ref{eq13}), $M_{Z}^{n}$ denotes the total scattering amplitude that is coming from the contribution of the $Z$-boson, while $M_{\gamma}^{n}$ is coming from the free $\gamma$-exchange, and they are given by:
\begin{eqnarray}
M_{Z}^{n}&=& \frac{e^{2}}{2C_{W}S_{W}} \frac{(1-2S_{W}^{2})}{S_{W}C_{W}} \frac{1}{(q_{1}+q_{2}+nk)^{2}-M_{Z}^{2}}  \Bigg\lbrace (p_{4}-p_{3})_{\mu}\bar{v}(p_{2}, s_{2}) \nonumber \\ &\times & \Bigg[ \kappa_{0}^{\mu}\,J_{n}(z)e^{-in\phi _{0}}(z)+ \frac{1}{2} \,\, \kappa_{1}^{\mu}\Big(J_{n+1}(z)e^{-i(n+1)\phi _{0}} + J_{n-1}(z)e^{-i(n-1)\phi _{0}}\Big) \nonumber \\ &+& \frac{1}{2\, i}\, \kappa_{2}^{\mu}\Big(J_{n+1}(z)e^{-i(n+1)\phi _{0}}-J_{n-1}(z)e^{-i(n-1)\phi _{0}}\Big)\Bigg] u(p_{1}, s_{1})  \Bigg\rbrace,
\label{eq14}
\end{eqnarray}
\begin{eqnarray}
M_{\gamma}^{n}&=& \frac{2^{}e^{2}}{(q_{1}+q_{2}+nk)^{2}} \Bigg\lbrace (p_{4}-p_{3})_{\mu}\bar{v}(p_{2}, s_{2})\Bigg[ \lambda_{0}^{\mu}\,J_{n}(z)e^{-in\phi _{0}}(z)\nonumber \\ &+ &\frac{1}{2} \,\, \lambda_{1}^{\mu}\Big(J_{n+1}(z)e^{-i(n+1)\phi _{0}} + J_{n-1}(z)e^{-i(n-1)\phi _{0}}\Big)\nonumber + \frac{1}{2\, i}\,\lambda_{2}^{\mu}\\ &\times& \Big(J_{n+1}(z)e^{-i(n+1)\phi _{0}}-J_{n-1}(z)e^{-i(n-1)\phi _{0}}\Big)\Bigg] u(p_{1}, s_{1}) \Bigg\rbrace.
\label{eq15}
\end{eqnarray}
$n$ is interpreted as the number of exchanged photons between the colliding physical system and the laser field. The six quantities that appear in equations \ref{eq14} and \ref{eq15} are related to Dirac matrices such that: 
\begin{equation}
\begin{cases}\kappa_{0}^{\mu}=\gamma^{\mu}(g_{v}^{e}-g_{a}^{e}\gamma^{5})+2b_{p_{1}}b_{p_{2}}\eta^{2}k^{\mu}\slashed k(g_{v}^{e}-g_{a}^{e}\gamma^{5})   &\\ \kappa_{1}^{\mu}=b_{p_{1}}\gamma^{\mu}(g_{v}^{e}-g_{a}^{e}\gamma^{5})\slashed k\slashed \eta_{1}-b_{p_{2}}\slashed \eta_{1}\slashed k \gamma^{\mu}(g_{v}^{e}-g_{a}^{e}\gamma^{5})   &\\ \kappa_{2}^{\mu}=b_{p_{1}}\gamma^{\mu}(g_{v}^{e}-g_{a}^{e}\gamma^{5})\slashed k\slashed \eta_{2}-b_{p_{2}}\slashed \eta_{2}\slashed k \gamma^{\mu}(g_{v}^{e}-g_{a}^{e}\gamma^{5}) 
  &\\\lambda_{0}^{\mu}=\gamma^{\mu}+2b_{p_{1}}b_{p_{2}}\eta^{2}k^{\mu}\slashed k
   &\\\lambda_{1}^{\mu}=b_{p_{1}}\gamma^{\mu}\slashed k\slashed \eta_{1}-b_{p_{2}}\slashed \eta_{1}\slashed k \gamma^{\mu}   
   &\\\lambda_{2}^{\mu}=b_{p_{1}}\gamma^{\mu}\slashed k\slashed \eta_{2}-b_{p_{2}}\slashed \eta_{2}\slashed k \gamma^{\mu},
\end{cases}
\label{eq16}
\end{equation}
where $b_{p_{i=1,2}}=e/2(k.p_{i})$. We have used the generating function of Bessel functions defined by:
\begin{equation}
e^{iz\sin\phi} = \sum_{n=-\infty}^{n=+\infty} J_{n}(z)e^{in\phi},
\label{eq17}
\end{equation}
with $z$ denotes the argument of the Bessel function and $\phi_0$ its phase. They are determined by:
\begin{equation}
z=\sqrt{\alpha_{1}^{2}+\alpha_{2}^{2}},\quad\qquad \phi_{0}=\arctan(\alpha_{1}/\alpha_2),
\label{eq18}
\end{equation}
where
\begin{equation}
\alpha_{1}=e\Bigg(\dfrac{(\eta_{1}.p_{1})}{(k.p_{1})}-\dfrac{(\eta_{1}.p_{2})}{(k.p_{2})}\Bigg), \quad \qquad \alpha_{2}=e\Bigg(\dfrac{(\eta_{2}.p_{1})}{(k.p_{1})}-\dfrac{(\eta_{2}.p_{2})}{(k.p_{2})}\Bigg).
\label{eq19}
\end{equation}
As it is known, to determine the differential cross section in the center of mass frame, we divide the squared matrix element by $VT$ and by the density of particles $\rho = V^{-1}$, then by the current of the incoming particles given by $|J_{inc}|=(\sqrt{(q_{1}q_{2})^{2}-m_{e}^{*^{4}}}/{Q_{1}Q_{2}V})$. We obtain the expression of the partial differential cross section as:
\begin{equation}
d\sigma_{n}=\dfrac{|S_{fi}^{n}|^{2}}{VT}\frac{1}{|J_{inc}|}\frac{1}{\varrho}V\int_{}\dfrac{d^{3}p_{3}}{(2\pi)^3}V\int_{}\dfrac{d^{3}p_{4}}{(2\pi)^3}.
\label{20}
\end{equation}
By averaging over the initial spins and summing over the final ones, and after some algebraic calculations, the differential cross section will be as follows:
\begin{eqnarray} 
d\bar{\sigma_{n}}({e}^{+}{e}^{-}\rightarrow H^{++}H^{--})&=&\dfrac{1}{16\sqrt{(q_{1}q_{2})^2-m_{e}^{*^{4}}}}   \big|\overline{M_{Z}^{n} + M_{\gamma}^{n}} \big|^{2} \int_{}\dfrac{|\mathbf{p}_{3}|^{2}d|\mathbf{p}_{3}|d\Omega}{(2\pi)^2E_{H^{--}}}\int_{}\dfrac{d^{3}p_{4}}{ E_{H^{++}}} \nonumber \\  &\times & \delta^{4}(p_{3}+p_{4}-q_{1}-q_{2}-nk).
\label{21}
\end{eqnarray}
The remaining integral over $d^{3}p_{4}$ can be evaluated by using the following formula \cite{35}:
\begin{equation}
 \int d\mathbf x f(\mathbf x) \delta(g(\mathbf x))=\dfrac{f(\mathbf x)}{|g^{'}(\mathbf x)|_{g(\mathbf x)=0}}.
\label{eq22}
\end{equation}
Finally, the differential cross section corresponding to the doubly charged Higgs bosons pair production via $e^{+}e^{-}$ annihilation is given by:
\begin{eqnarray}
\dfrac{d\sigma_{n}}{d\Omega}({e}^{+}{e}^{-}\rightarrow H^{++}H^{--})&=&\dfrac{1}{16\sqrt{(q_{-}q_{+})^2-m_{e}^{*^{4}}}}   \big|\overline{M_{Z}^{n} + M_{\gamma}^{n}} \big|^{2} \dfrac{2|\mathbf{p}_{3}|^{2}}{(2\pi)^2E_{H^{--}}} \nonumber \\  &\times & \dfrac{1}{|g^{'}(|\mathbf{p}_{3}|)|_{g(|\mathbf{p}_{3}|)=0}},
\label{23}
\end{eqnarray}
where
\begin{equation}
\big|g^{'}(|\textbf{p}_{3}|)\big|=-4\dfrac{|\textbf p_{3}|}{\sqrt{|\textbf p_{3}|^{2}+M_{H^{\pm\pm}}^{2}}}\Bigg[\dfrac{(\sqrt{s}+n\omega)}{2}-\frac{e^{2}a^{2}}{\sqrt{s}}\Bigg],
\label{24}
\end{equation}
with $M_{H^{\pm\pm}}$ is the mass of the doubly charged Higgs $H^{\pm\pm}$. To find the expression of the quantity $\big|\overline{M_{\gamma}^{n} + M_{Z}^{n}} \big|^{2}$ given in equation \ref{23}, we have used the FeynCalc program \cite{36}, and its expression is given in the appendix.
We have checked the gauge invariance of the total cross section, and for simplicity reasons, we have considered the Feynman gauge, $\xi=1$, in our calculations.
\section{Results and discussion}\label{sec3}
In this section, we present and analyze our finding results for the doubly charged Higgs boson production process through $e^{+}e^{-}$ annihilation in the centre of mass frame. Our calculations focus on the case in which the polarization vector of the laser field is taken to be parallel to the $z$-axis. Moreover, we can obtain the total cross section by performing a numerical integration of the differential cross section, given by equation (\ref{23}), over the solid angle $d\Omega$. 
For the numerical analysis, we take the SM parameters from PDG \cite{37} such that $m_{e}=0.511\, MeV$, $M_{Z}=91.1875\, GeV$, and the mixing angle $\sin^{2}\theta_W = 0.23126$. 
As we have shown in equation \ref{23}, besides the mass of the doubly charged Higgs boson $M_{H^{\pm\pm}}$ and $e^{+}e^{-}$ colliding energy, the cross section depends on the electromagnetic field parameters such as the number of exchanged photons $n$, the laser field strength $\varepsilon_{0}$ and its frequency $\omega$. 
Before presenting the results of our scan, we want to test our numerical calculations for the obtained laser assisted total cross section. Therefore, we have chosen to compare our numerical results with the results obtained in \cite{33}. To do this, we have taken the following conditions: The laser field strength is equal to zero, and we consider that no photon is transferred between the physical system and the laser field ($n=0$).
\begin{figure}[H]
  \centering
      \includegraphics[scale=0.6]{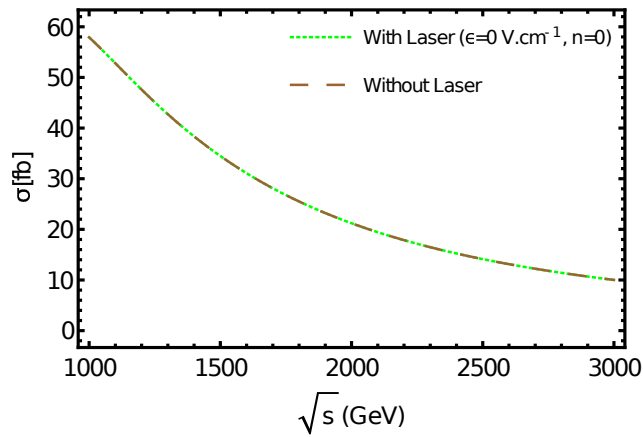}
  \caption{The total cross section of the process $e^{+}e^{-}\rightarrow H^{++}H^{--}$ as a function of $\sqrt{s}$ with and without laser field by choosing the laser parameters as $\varepsilon_{0} = 0\, V.cm^{-1}$ and $ n = 0$, and $M_{H}$ is taken as $300\, GeV$.}
  \label{fig2}
\end{figure}
In figure \ref{fig2}, we compare the laser-assisted total cross section of the process $e^{+}e^{-}\rightarrow H^{++}H^{--}$ in the presence of a circularly polarized laser field with its corresponding laser-free total cross section \cite{33}. The presence of a laser field implies a long and complicated calculation. This makes it difficult for us to verify our calculations analytically, and this comparison technique provides the possibility of testing the validity of our results.
We clearly observe from figure \ref{fig2} that, if we consider the laser field parameters as zero, the two total cross sections are very similar, and this result is in excellent agreement with our theoretical calculation. 
Now, we will focus our analysis on the effect of a circularly polarized electromagnetic field on the scattering process \ref{fig1}. We start our discussion with a very important point which concerns the behavior of the partial total cross section, that corresponds to each four-momentum conservation $\delta(p_{3}+p_{4}-q_{1}-q_{2}-nk) = 1$, versus the number of exchanged photons.
\begin{table}[H]
 \centering
\caption{\label{tab1}Laser-assisted total cross section as a function of the number of exchanged photons for different laser field strengths and frequencies. The centre of mass energy and the doubly charged Higgs mass are chosen as $\sqrt{s}=1000\,GeV$ and $M_{H^{\pm\pm}}=300\,GeV$, respectively.}
\begin{tabular}{ccccccccc}
\hline
  & & $ \sigma $[fb]  & &$ \sigma $[fb] & & $ \sigma $[fb] &  \\
 $ \varepsilon_{0}(\,V.cm^{-1}) $ && $ CO_{2} $ Laser  & & Nd:YAG laser & & He:Ne Laser  &\\
& n &  $\omega=0.117\,eV $ & n &$\omega=1.17\,eV $& n & $\omega=2\,eV  $ & \\
 \hline
 
 &$\pm1300$ & $ 0 $ & $\pm18$ & $ 0 $ & $\pm8$ & $ 0 $ \\
   &$\pm1050$ & $ 0 $ & $\pm15$ & $ 0 $ & $\pm6$ & $ 0 $ \\
  $ 10^{5} $  &$\pm900$ & $ 0.00043251 $ & $\pm12$ & $ 0.319588 $ & $\pm4$ & $ 2.41077 $ \\
   &$\pm600$ & $ 0.0426101 $ & $\pm8$ & $ 5.45433 $ & $\pm3$ & $ 8.64928 $ \\
    &$\pm300$ & $ 0.0268137 $ & $\pm4$ & $ 1.95167 $ & $\pm2$ & $ 12.1979 $ \\
    &$0$ & $ 0.0347302  $ & $0$ & $ 3.60327 $  & $0$ & $ 8.35336 $ \\
     \hline
 &$\pm6000$ & $ 0 $ & $\pm150$ & $ 0 $ & $\pm50$ & $ 0 $ \\
   &$\pm5100$ & $ 0 $ & $\pm120$ & $ 0 $ & $\pm40$ & $ 0 $ \\
  $ 10^{6} $  &$\pm4000$ & $ 0.000585962 $ & $\pm90$ & $ 0.756126 $ & $\pm30$ & $ 0.660953 $ \\
   &$\pm2000$ & $ 0.000651189 $ & $\pm60$ & $ 0.425914 $ & $\pm20$ & $ 0.666273 $ \\
   &$\pm1000$ & $ 0.000741295 $ & $\pm30$ & $ 0.122119 $ & $\pm10$ & $ 0.255696 $ \\
    &$0$ & $ 0.0030297  $ & $0$ & $ 0.145686 $  & $0$ & $ 0.915362 $ \\
     \hline
 &$\pm12000$ & $ 0 $ & $\pm1300$ & $ 0 $ & $\pm500$ & $ 0 $ \\
   &$\pm9150$ & $ 0 $ & $\pm1100$ & $ 0 $ & $\pm400$ & $ 0 $ \\
  $ 10^{7} $  &$\pm6000$ & $ 0.000360217 $ & $\pm900$ & $ 0.000432513 $ & $\pm300$ & $ 0.0122968 $ \\
   &$\pm3000$ & $ 0.000294922 $ & $\pm600$ & $ 0.0426101 $ & $\pm200$ & $ 0.0969274 $ \\
   &$\pm1000$ & $ 0.000268565 $ & $\pm300$ & $ 0.0268137 $ & $\pm100$ & $ 0.0261197 $ \\
    &$0$ & $ 0.000347827  $ & $0$ & $ 0.0347302 $  & $0$ & $ 0.0989345 $ \\
     \hline
\end{tabular}
\end{table}

\begin{figure}[H]
  \centering
      \includegraphics[scale=0.6]{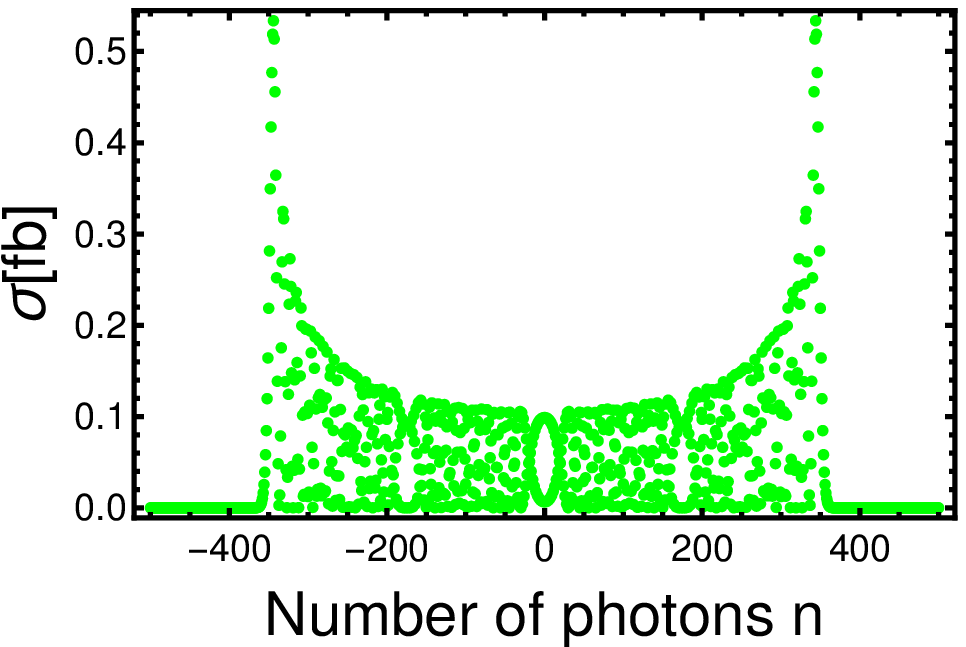}\hspace*{0.4cm}
      \includegraphics[scale=0.6]{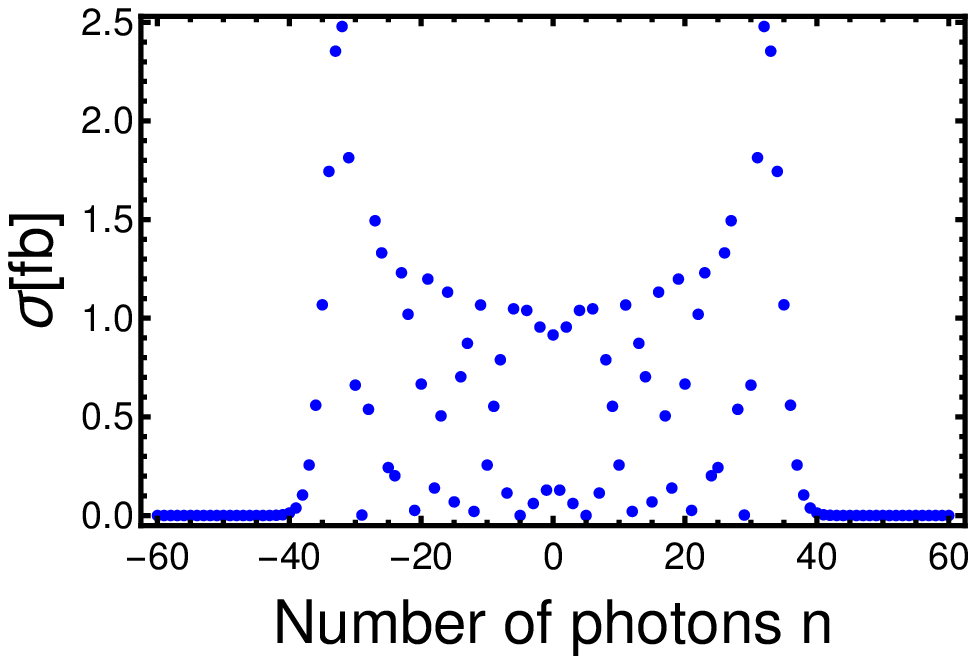}\par\vspace*{0.5cm}
        \caption{Laser-assisted total cross section as a function of the number of exchanged photons by taking the centre of mass energy and the charged Higgs mass as $\sqrt{s}=1000\,GeV$ and $M_{H^{\pm\pm}}=300\,GeV$, respectively. The  \textbf{He:Ne Laser} $(\omega=2\,eV)$ is used in both figures with $\varepsilon_{0}=10^{7}\,V.cm^{-1}$ (left panel) and $\varepsilon_{0}=10^{6}\,V.cm^{-1}$ (right panel).}
        \label{fig3}       
\end{figure}

We illustrate, in table \ref{tab1}, the variation of the partial total cross section of the process $e^{+}e^{-}\rightarrow H^{++}H^{--}$ versus the number of exchanged photons $n$. This variation is presented for different laser field strengths and frequencies. 
The transfer of photons between the laser field and the scattering process indicates that the incoming particles are interacting with the laser field. In addition, we see that for the laser field strength, $\varepsilon_{0}=10^{5}\, V.cm^{-1}$, and for a specific laser frequency $\omega=0.117\, eV$, a significant number of photons can be exchanged ($\pm1050$ photons) between the laser field and the colliding physical system. Moreover, this number of exchanged photons is enhanced with the increase of the laser field strength.
For instance, for $\varepsilon_{0}=10^{6}\, V.cm^{-1}$, the greatest value of photons number that can be exchanged is equal to $\pm5100$, while the cutoff number is $\pm9150$ for the case where $\varepsilon_{0}=10^{7}\, V.cm^{-1}$. This result indicates that the incident particles interact strongly with high laser field strengths. 
Thus, the effect of electromagnetic field on the electron positron scattering process becomes prominent and important, and, as a consequence, the partial total cross section will be affected and changed.
As an example, figure \ref{fig3} shows the partial cross section as a function of the number of laser photons absorbed ($n > 0$) or emitted ($n < 0$) for $\omega=2\, eV$ and for two different known laser strengths. One can see in figure \ref{fig3} (left panel) that the maximum number of photons that can be transferred and for which the partial total cross section vanishes for $\varepsilon_{0}=10^{7}\, V.cm^{-1}$ is greater than that corresponds to $\varepsilon_{0}=10^{6}\, V.cm^{-1}$ (right panel).  Therefore, these results are in good agreement with those given in table \ref{tab1}.
\begin{figure}[H]
  \centering
      \includegraphics[scale=0.62]{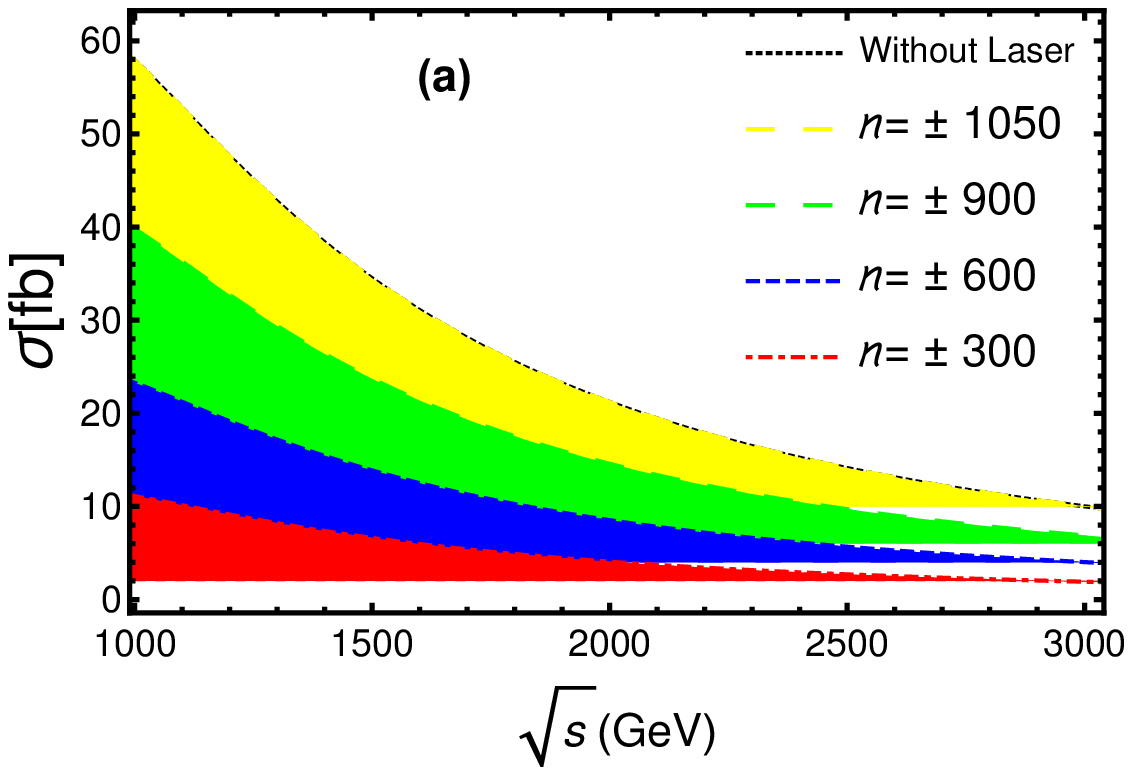}\hspace*{0.4cm}
      \includegraphics[scale=0.62]{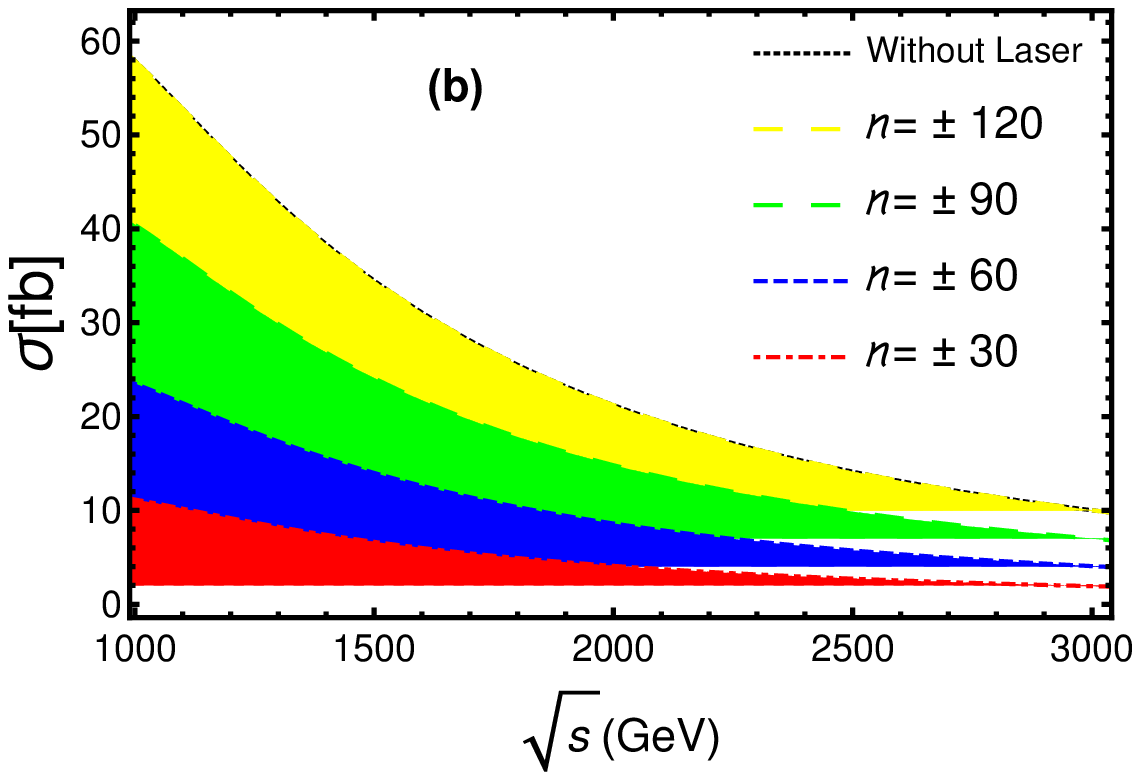}\par\vspace*{0.5cm}
      \includegraphics[scale=0.64]{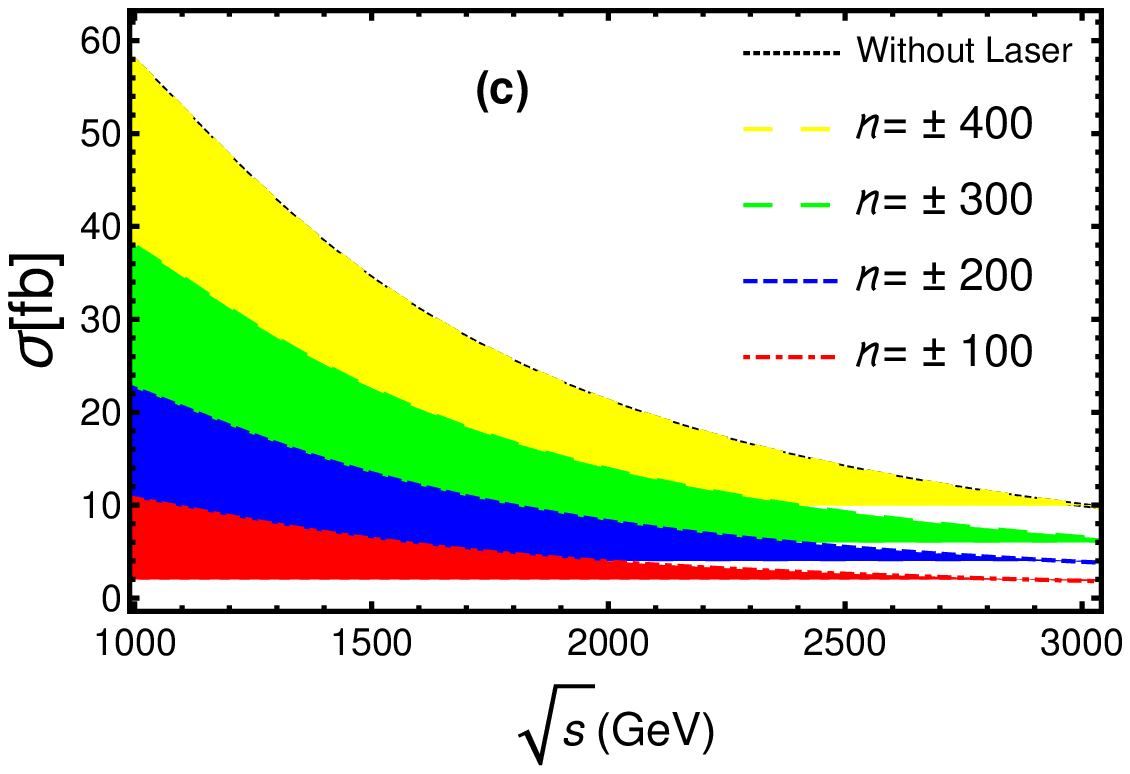}
        \caption{The laser-assisted total cross section of $e^{+}e^{-}\rightarrow H^{++}H^{--}$ as a function of the centre of mass energy for different exchanged photons number and by taking $M_{H^{\pm\pm}}=300\,GeV$ . The laser field strength and its frequency are chosen as:  $\epsilon_{0}=10^{5}V.cm^{-1}$ and $(\omega=0.117\, eV)$ in (\textbf{a});  $\epsilon_{0}=10^{6}V.cm^{-1}$ and  $(\omega=1.17\, eV)$ in (\textbf{b});  $\epsilon_{0}=10^{7}V.cm^{-1}$ and  $(\omega=2\, eV)$ in (\textbf{c}).}
        \label{fig4}
\end{figure}
In figure \ref{fig4}, we plot the dependence of the laser-assisted total cross section $\sigma$ on the centre of mass energy for different laser parameters. We use different colors to clearly show the influence of the laser field on the order of magnitude of the total cross section. As it can be seen, due to the phase space suppression, the total cross section $\sigma$ declines by increasing $\sqrt{s}$. 
Moreover, for all cases, the exchange of a large number of photons will always give a quite large values of cross section until $n$ reaches $\pm$ cutoff.
For instance, in figure \ref{fig4}(\textbf{b}), in which $\varepsilon_{0}=10^{6}\, V.cm^{-1}$ and $\omega=1.17\, eV$, the total cross section remains under $24fb$ for $n=\pm 60$, while it reaches up to $40fb$ for $n=\pm 90$. The summation over $\pm$ cutoff number is called sum-rule \cite{38}, and it leads to a cross section which is equal to its corresponding laser-free cross section in all centre of mass energies. Another important point which should be discussed, here, is that the laser parameters have a great effect on the order of magnitude of the cross section. 
By comparing figures \ref{fig4}(\textbf{a}) and \ref{fig4}(\textbf{c}), for the same number of exchanged photons such as $n=\pm 300$, we observe that the maximum value of the cross section mostly remains under $12fb$ for $\varepsilon_{0}=10^{5}\, V.cm^{-1}$ and $\omega=0.117\, eV$, and it increases to about $38fb$ for $\varepsilon_{0}=10^{7}\, V.cm^{-1}$ and $\omega=2\, eV$.
\begin{figure}[H]
  \centering
      \includegraphics[scale=0.7]{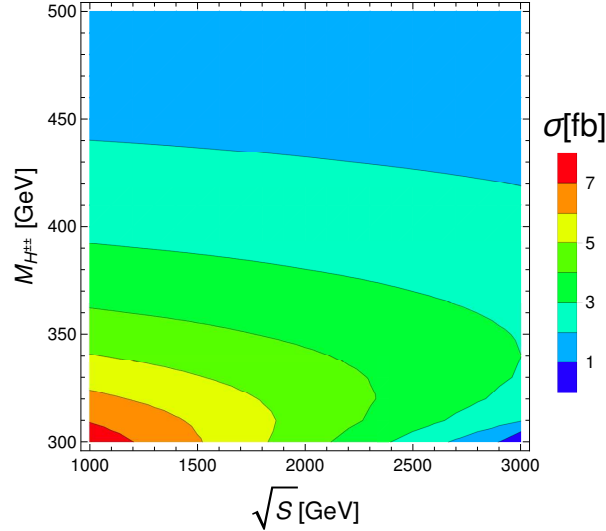}
  \caption{The laser-assisted total cross section of the process $e^{+}e^{-}\rightarrow H^{++}H^{--}$ as a function of the doubly charged Higgs mass and $e^{+}e^{-}$ colliding energy by summing over $n$ from $-20$ to $20$ and by taking the laser field strength and its frequency as $\varepsilon_{0}=10^{6}\, V.cm^{-1}$ and $\omega=1.17\, eV$, respectively.}
  \label{fig5}
\end{figure}
\begin{table}[H]
 \centering
 \caption{\label{tab2}Laser-assisted total cross section of $H^{++}H^{--}$ production at $\sqrt{s}=1000\, GeV$  for some typical values of $M_{H^{\pm\pm}}$. The laser's parameters and the number of exchanged photons are chosen as:  $\varepsilon_{0}=10^{6}\, V.cm^{-1}$, $\omega=1.17\, eV$ and $n=\pm 20$.}
\begin{tabular}{cccccccccccc}
        \hline
        \hline
         $M_{H^{\pm\pm}}\, [GeV]$  & 320 & 320 & 360 & 380 & 400 & 420 & 440 & 460 & 480 & 500 \\ 
        \hline
        $\sigma\, [fb]$ & 7.021 & 6.399 & 5.754 & 5.089 & 4.407 & 3.709 & 2.992 & 2.249 & 1.443 & 1.132 \\ 
        \hline
        \hline
\end{tabular}
\end{table}
In order to study the dependence of the total cross section $\sigma$ on the centre of mass energy and the mass of doubly charged Higgs, we present in figure \ref{fig5} the total laser-assisted cross section in the ($M_{H^{\pm\pm}}$,$\sqrt{s}$) plane. The numerical results for some typical values of $M_{H^{\pm\pm}}$ are also given in table \ref{tab2} where $\sqrt{s}=1000\, GeV$. From figure \ref{fig5}, we can see that for light doubly charged Higgs mass, the total cross section is rather important. This imposes a sever constraint on $\sqrt{s}$. 
Namely, for $M_{H^{\pm\pm}}<310\, GeV$, $\sqrt{s}$ is obliged to be in the range of $1000\sim 1200\, GeV$. 
The range of $\sqrt{s}$ becomes more large as long as the doubly charged Higgs mass increases. Consequently, the total cross section declines. 
It is remarkable that there exists a small region of light doubly charged Higgs boson, i.e. the doubly charged Higgs boson mass can be around $300\, GeV$, and with a large $\sqrt{s}$ ($\sqrt{s}>2900\, GeV$) in which the total cross section is less than $1fb$. 
In table \ref{tab2}, it is clear that the total laser-assisted cross section decreases from about $7.02fb$ to around $1.13fb$ as $M_{H^{\pm\pm}}$ increases from $320\, GeV$ to $500\, GeV$. 
We mention that the chosen doubly charged Higgs mass in table \ref{tab2} are consistent with the HTM constraints \cite{33}.
\begin{figure}
  \centering
      \includegraphics[scale=0.58]{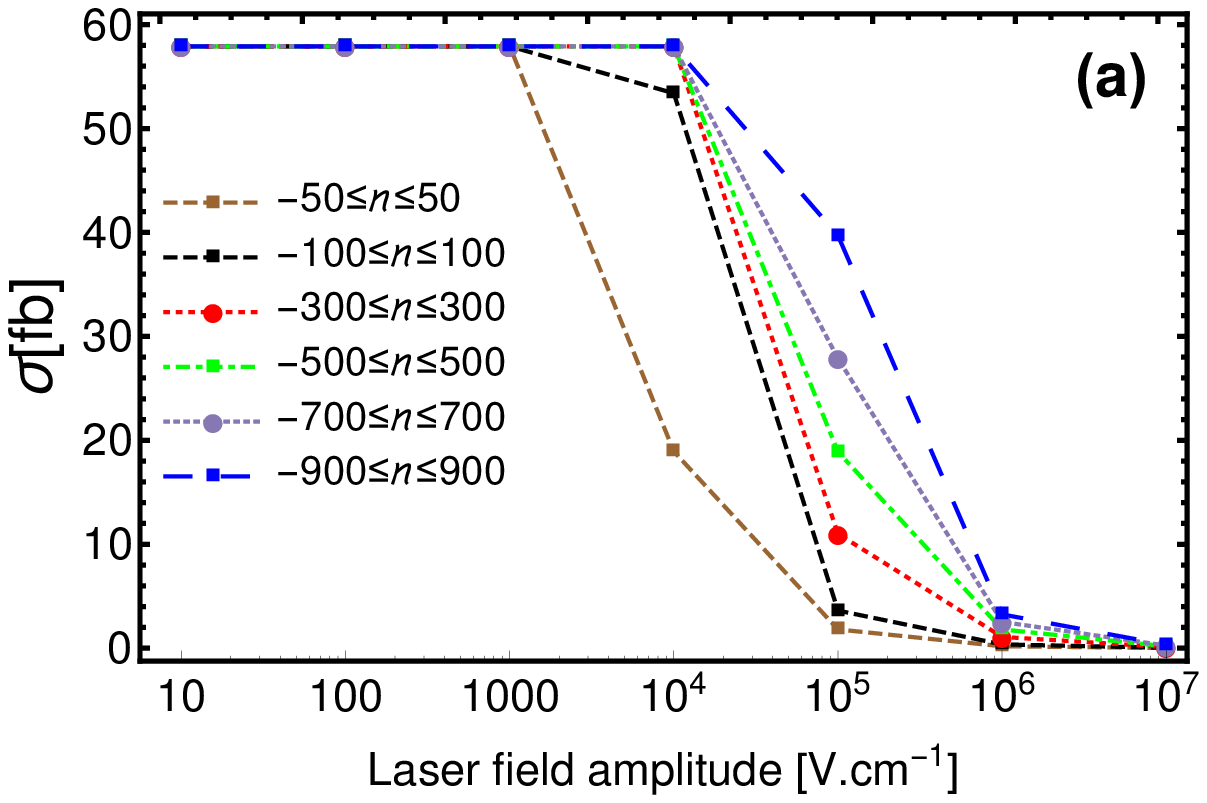}\hspace*{0.4cm}
      \includegraphics[scale=0.58]{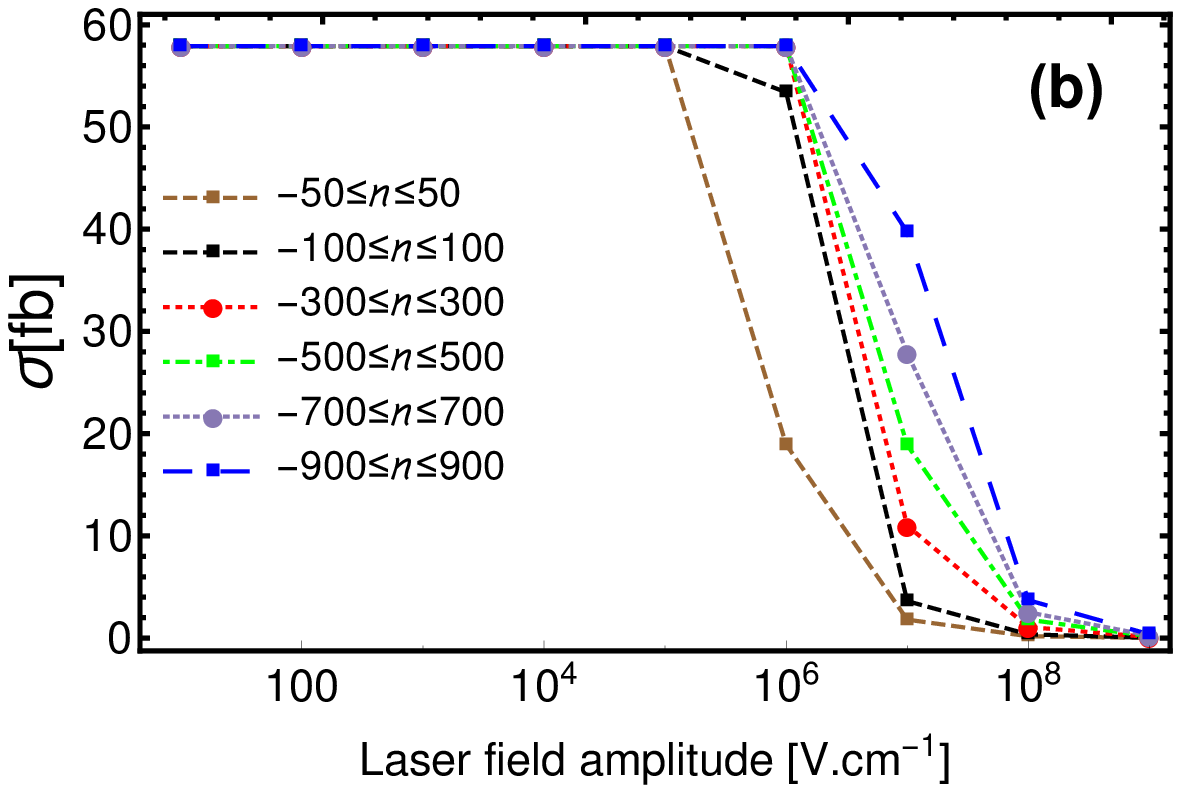}\par\vspace*{0.5cm}
      \includegraphics[scale=0.7]{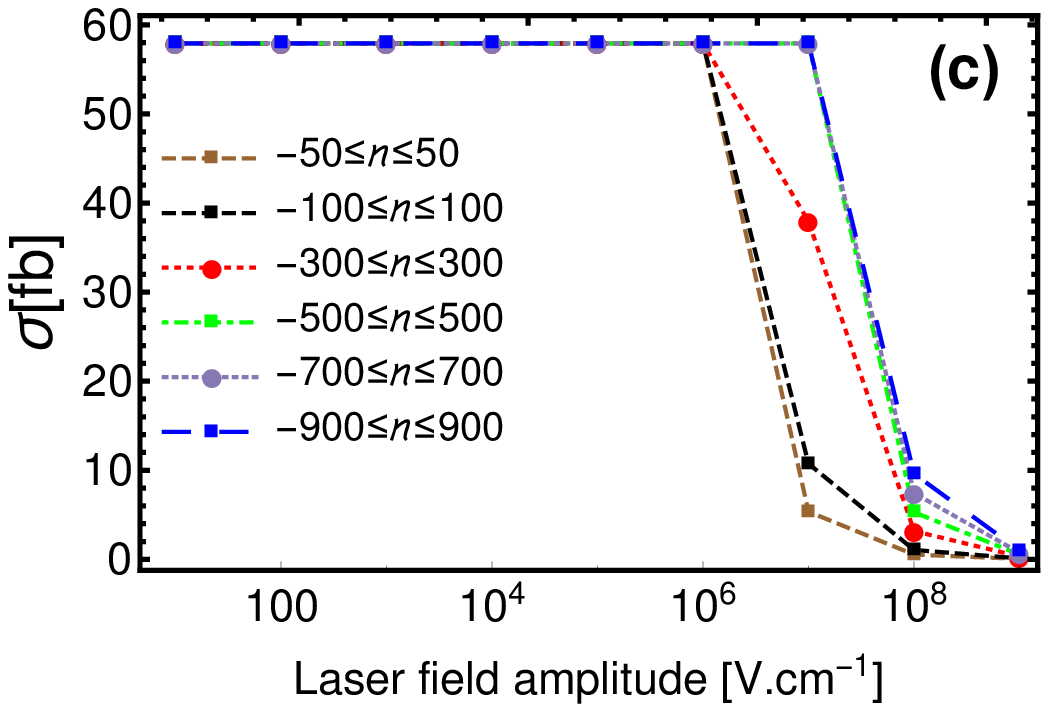}
        \caption{Dependence of the laser-assisted total cross section on the laser field strength for different exchanged numbers of photons. The centre of mass energy is taken as $\sqrt{s}=1000\, GeV$ for all curves. The laser frequency in (\textbf{a}), (\textbf{b}) and (\textbf{c}) is $\,\omega=0.117 eV$, $\omega=1,17 eV$ and $\omega=2 eV$, respectively.}
      \label{fig6}
\end{figure}
Before ending this section, we present in figure \ref{fig6} the influence of the laser field amplitude on the cross section for different known frequencies and different number of exchanged photons which are indicated in each plot by the same type of lines. From this plot, it is obvious that, for small laser intensities $\varepsilon_{0}<10^{3}\, V.cm^{-1}$, all curves are very similar. This means that the impact of laser field on the incoming particles is strongly suppressed. Consequently, no photon will be transferred between the laser field and the colliding physical system. 
From figure \ref{fig6}(\textbf{a}), in which $\omega=0.117\, eV$, both curves which correspond to $n=\pm 50$ and $n=\pm 100$ begin to deviate from $\varepsilon_{0}\geq 10^{3}\, V.cm^{-1}$, while the other curves begin to deviate from each other when $\varepsilon_{0}\geq 10^{4}\, V.cm^{-1}$. 
Therefore, the laser-assisted total cross section decreases until it becomes zero. Moreover, by comparing figures \ref{fig6}(\textbf{a}) and \ref{fig6}(\textbf{b}), we remark that, for the same number of exchanged photons such as $n=\pm 900$, the order of magnitude of the total cross-section vanishes as much as the laser field strength is very close to $10^{7}\, V.cm^{-1}$ and $10^{9}\, V.cm^{-1}$ for $\omega=0.117\, eV$ and $\omega=1.17\, eV$, respectively.
\section{Conclusion}\label{concl}
Searching for new Higgs boson state may reveal first signs of new physics beyond the standard model (BSM). In this work, we study in detail the doubly charged Higgs pair production at future electron-positron colliders in the presence of a circularly polarized electromagnetic field within the framework of HTM. We calculate the total cross section of the process $e^{+}e^{-}\rightarrow H^{++}H^{--}$ inside the laser field.
Then, we study the dependence of the production cross section on $M_{H^{\pm\pm}}$ and $e^{+}e^{-}$ colliding energy as well as on the laser field parameters such as the number of exchanged photons, the laser field strength and its frequency.
The numerical results show that there is a correlation between the number of exchanged photons and the laser field strength. More accurately, this number enhances as long as the laser field strength increases. 
In addition, as far as the number $n$ takes high values, the total laser-assisted cross section increases until it becomes equal to its corresponding laser-free cross section. 
We have provided some numerical results for $\sigma$ as a function of $M_{H^{\pm\pm}}$ for different known frequencies at $\sqrt{s}=1000\, GeV$. We have found that the total cross section varies in the range of $7\sim 1.13\, fb$ with the increment of $M_{H^{\pm\pm}}$ from $300$ to $500\,GeV$. Finally, we indicate that the behavior of the total cross section doesn't change for small laser intensities while it is affected and changed for high laser field strengths.
\section{Appendix}
In this appendix, we give the expression of the quantity $\big|\overline{M_{\gamma}^{n} + M_{Z}^{n}} \big|^{2}$ that appears in equation (\ref{23}).
\scriptsize
\begin{eqnarray}
 \big|\overline{M_{\gamma}^{n} + M_{Z}^{n}} \big|^{2}&=&\nonumber\dfrac{1}{4}\sum_{n=-\infty}^{+\infty}\sum_{s}\big| M_{\gamma}^{n} + M_{Z}^{n} \big|^{2}=\frac{1}{4}\sum_{n=-\infty}^{+\infty}\Bigg\lbrace\dfrac{4e^{4}}{(q_{1}+q_{2}+nk)^{4}} Tr\Bigg[(p_{4_{\mu}}-p_{3_{\mu}})(\slashed p_{1}-m_{e})\\ \nonumber &\times &\Big[ \lambda_{0}^{\mu}\,J_{n}(z)e^{-in\phi _{0}}(z)+\lambda_{1}^{\mu}\,\,\frac{1}{2}\Big(J_{n+1}(z)e^{-i(n+1)\phi _{0}} + J_{n-1}(z)e^{-i(n-1)\phi _{0}}\Big) \\ &+ & \nonumber\lambda_{2}^{\mu}\,\frac{1}{2\, i}\Big(J_{n+1}(z)e^{-i(n+1)\phi _{0}}-J_{n-1}(z)e^{-i(n-1)\phi _{0}}\Big)\Big](p_{4_{\nu}}-p_{3_{\nu}})(\slashed p_{2}+m_{e})\\ \nonumber &\times &  \Big[ \lambda_{0}^{\nu}\,J^{*}_{n}(z)e^{+in\phi _{0}}(z)+\lambda_{1}^{\nu}\,\,\frac{1}{2}\Big(J^{*}_{n+1}(z)e^{+i(n+1)\phi _{0}} + J^{*}_{n-1}(z)e^{+i(n-1)\phi _{0}}\Big) \\ &- & \nonumber\lambda_{2}^{\nu}\,\frac{1}{2\, i}\Big(J^{*}_{n+1}(z)e^{+i(n+1)\phi _{0}}-J^{*}_{n-1}(z)e^{+i(n-1)\phi _{0}}\Big)\Big]\Bigg]
+\left(\dfrac{e^{2}}{2C_{W}S_{W}}\right)^{2} \left(\frac{(1-2S_{W}^{2})}{S_{W}C_{W}}\right)^{2}\\ \nonumber &\times &  \left(\frac{1}{(q_{1}+q_{2}+nk)^{2}-M_{Z}^{2}}\right)^{2} Tr\Bigg[(p_{4_{\mu}}-p_{3_{\mu}})(\slashed p_{1}-m_{e})\Big[ \kappa_{0}^{\mu}\,J_{n}(z)e^{-in\phi _{0}}(z)\\ \nonumber &+ &\kappa_{1}^{\mu}\,\,\frac{1}{2}\Big(J_{n+1}(z)e^{-i(n+1)\phi _{0}} + J_{n-1}(z)e^{-i(n-1)\phi _{0}}\Big)+\kappa_{2}^{\mu}\,\frac{1}{2\, i}\Big(J_{n+1}(z)e^{-i(n+1)\phi _{0}}-J_{n-1}(z)e^{-i(n-1)\phi _{0}}\Big)\Big]\\ \nonumber &\times & (p_{4_{\nu}}-p_{3_{\nu}})(\slashed p_{2}+m_{e}) \Big[ \kappa_{0}^{\nu}\,J^{*}_{n}(z)e^{+in\phi _{0}}(z)+ \kappa_{1}^{\nu}\,\,\frac{1}{2}\Big(J^{*}_{n+1}(z)e^{+i(n+1)\phi _{0}} + J^{*}_{n-1}(z)e^{+i(n-1)\phi _{0}}\Big) \\ &- & \nonumber\kappa_{2}^{\nu}\,\frac{1}{2\, i}\Big(J^{*}_{n+1}(z)e^{+i(n+1)\phi _{0}}-J^{*}_{n-1}(z)e^{+i(n-1)\phi _{0}}\Big)\Big]\Bigg]
+ \dfrac{2e^{2}}{(q_{1}+q_{2}+nk)^{2}}\left(\dfrac{e^{2}}{2C_{W}S_{W}}\right)\\ \nonumber &\times & \left(\frac{(1-2S_{W}^{2})}{S_{W}C_{W}}\right)  \frac{1}{(q_{1}+q_{2}+nk)^{2}-M_{Z}^{2}} Tr\Bigg[(p_{4_{\mu}}-p_{3_{\mu}})(\slashed p_{1}-m_{e})\Big[ \lambda_{0}^{\mu}\,J_{n}(z)e^{-in\phi _{0}}(z) \\ &+ &\nonumber\lambda_{1}^{\mu}\,\,\frac{1}{2}\Big(J_{n+1}(z)e^{-i(n+1)\phi _{0}} + J_{n-1}(z)e^{-i(n-1)\phi _{0}}\Big)+ \lambda_{2}^{\mu}\,\frac{1}{2\, i}\Big(J_{n+1}(z)e^{-i(n+1)\phi _{0}}-J_{n-1}(z)e^{-i(n-1)\phi _{0}}\Big)\Big]\\ \nonumber &\times & (p_{4_{\nu}}-p_{3_{\nu}})(\slashed p_{2}+m_{e}) \Big[ \kappa_{0}^{\nu}\,J^{*}_{n}(z)e^{+in\phi _{0}}(z)+\kappa_{1}^{\nu}\,\,\frac{1}{2}\Big(J^{*}_{n+1}(z)e^{+i(n+1)\phi _{0}} + J^{*}_{n-1}(z)e^{+i(n-1)\phi _{0}}\Big)\\ \nonumber &- & \kappa_{2}^{\nu}\,\frac{1}{2\, i}\Big(J^{*}_{n+1}(z)e^{+i(n+1)\phi _{0}}-J^{*}_{n-1}(z)e^{+i(n-1)\phi _{0}}\Big)\Big]\Bigg]
+\dfrac{2e^{2}}{(q_{1}+q_{2}+nk)^{2}}\left(\dfrac{e^{2}}{2C_{W}S_{W}}\right)\\ \nonumber &\times & \left(\frac{(1-2S_{W}^{2})}{S_{W}C_{W}}\right) \frac{1}{(q_{1}+q_{2}+nk)^{2}-M_{Z}^{2}} Tr\Bigg[(p_{4_{\mu}}-p_{3_{\mu}})(\slashed p_{1}-m_{e})\\ \nonumber &\times & \Big[ \kappa_{0}^{\mu}\,J_{n}(z)e^{-in\phi _{0}}(z)+\kappa_{1}^{\mu}\,\,\frac{1}{2}\Big(J_{n+1}(z)e^{-i(n+1)\phi _{0}} + J_{n-1}(z)e^{-i(n-1)\phi _{0}}\Big)\\ \nonumber &+ & \kappa_{2}^{\mu}\,\frac{1}{2\, i}\Big(J_{n+1}(z)e^{-i(n+1)\phi _{0}}-J_{n-1}(z)e^{-i(n-1)\phi _{0}}\Big)\Big] (p_{4_{\nu}}-p_{3_{\nu}})(\slashed p_{2}+m_{e}) \Big[ \lambda_{0}^{\nu}\,J^{*}_{n}(z)e^{+in\phi _{0}}(z)\\ &+ & \nonumber\lambda_{1}^{\nu}\,\,\frac{1}{2}\Big(J^{*}_{n+1}(z)e^{+i(n+1)\phi _{0}} + J^{*}_{n-1}(z)e^{+i(n-1)\phi _{0}}\Big) -\lambda_{2}^{\nu}\,\frac{1}{2\, i}\Big(J^{*}_{n+1}(z)e^{+i(n+1)\phi _{0}}-J^{*}_{n-1}(z)e^{+i(n-1)\phi _{0}}\Big)\Big]\Bigg]\Bigg\rbrace.
\end{eqnarray}
       
\end{document}